\documentclass[conference]{IEEEtran}
\IEEEoverridecommandlockouts
\usepackage{cite}
\usepackage{amsmath,amssymb,amsfonts}
\usepackage{algorithmic}
\usepackage{graphicx}
\usepackage{textcomp}
\usepackage{xcolor}
\usepackage{adjustbox}
\usepackage[numbers]{natbib}
\usepackage{comment}
\usepackage{subcaption}
\usepackage{url}

\def\BibTeX{{\rm B\kern-.05em{\sc i\kern-.025em b}\kern-.08em
    T\kern-.1667em\lower.7ex\hbox{E}\kern-.125emX}}
\begin{document}

\title{Baba is Y'all:\\Collaborative Mixed-Initiative Level Design}

\author{\IEEEauthorblockN{Megan Charity}
\IEEEauthorblockA{\textit{Game Innovation Lab} \\
\textit{New York Univeristy}\\
Brooklyn, USA \\
mlc761@nyu.edu}
\and
\IEEEauthorblockN{Ahmed Khalifa}
\IEEEauthorblockA{\textit{Game Innovation Lab} \\
\textit{New York Univeristy}\\
Brooklyn, USA \\
ahmed@akhalifa.com}
\and
\IEEEauthorblockN{Julian Togelius}
\IEEEauthorblockA{\textit{Game Innovation Lab} \\
\textit{New York Univeristy}\\
Brooklyn, USA \\
julian@togelius.com}
}
\IEEEoverridecommandlockouts
\IEEEpubid{\makebox[\columnwidth]{978-1-5386-5541-2/18/\$31.00~\copyright2018 IEEE \hfill} \hspace{\columnsep}\makebox[\columnwidth]{ }}
\maketitle
\IEEEpubidadjcol
\begin{abstract}
We present a collaborative mixed-initiative system for building levels for the puzzle game ``Baba is You''. Unlike previous mixed-initiative systems, Baba is Y'all is designed for collaborative asynchronous creation by multiple users over the internet. The system includes several AI-assisted features to help designers, including a level evolver and an automated player for playtesting. The level archives catalogues levels according to which mechanics are implemented and not implemented, allowing the system to ask users to design levels with specific combinations of mechanics. We describe the operation of the system and the results of small-scale informal user test, and discuss future development paths for this system as well as for collaborative mixed-initiative systems in general.
\end{abstract}


\section{Introduction}

How to best design game content together with content generation algorithms is a hard and important question. A number of prototype systems for  \emph{mixed-initiative design} have been created to showcase ways in which humans and algorithms can design game content together~\cite{smith2010tanagra,liapis2013sentient,shaker2013ropossum}. Many different modes of interaction have been devised, including those where the computer program provides suggestions to the human designer, evaluates their output, tests for playability, etc. However, all of these systems for AI-assisted game content generation are geared towards a single user.

In this paper, we address the challenge of AI-assisted \emph{collaborative} game content creation, that is, where multiple users interact with a procedural content generation system to create game content. Similar to an open source system like Wikipedia, there would be a central content repository, where anyone could make a contribution to the content. But additionally, the system should help users create content through various AI functionalities, such as providing suggestions, testing, and feedback. Most importantly, the initiative for design should be mixed. For example, the system might ask users to design specific types of content that it thinks should be designed, or to test or evaluate artifacts that others might have designed.

In this paper, we describe a prototype system for collaborative mixed-initiative level design, where users design levels for the puzzle game Baba is You (Arvi Teikari, 2019). The system includes features for editing levels, automatically playtesting levels, helping design levels through an evolutionary algorithm, rating levels, and suggesting novel levels to design. All features are built around a central level archive which is structured like the map of elites from the MAP-Elites algorithm. We also report preliminary results from an informal user study, shining some light on how the system can be used.

\section{Background}

\subsection{Procedural Content Generation}
Procedural Content Generation is the process of using a computer program to create content~\cite{shaker2016procedural}. These techniques have been used since the early days of computer games and still remain a popular technique today. PCG is typically divided based on the technology behind the generation process into three main categories: Constructive techniques~\cite{shaker2016procedural}, Search-Based techniques~\cite{togelius2011search}, and Machine Learning techniques~\cite{summerville2018procedural}. Search based approaches are more common in academia for their generality and ease of use. These techniques use a search/optimization algorithm to search the space of potential content for good artifacts. A fitness function is defined in order to guide the search. Previously, researchers have used it to generate decorations such as flowers~\cite{risi2015petalz}, game mechanics such as bullet patterns~\cite{hastings2009automatic}, or entire game levels~\cite{khalifa2018talakat,liapis2013sentient}, etc.

With the advancement in quality-diversity search based methods~\cite{pugh2016quality}, game researchers have started to focus on using it in more projects~\cite{gravina2019procedural}. Quality-diversity techniques are search based techniques that try to generate a set of diverse solutions with a high quality. A well-known example is Map-Elites~\cite{mouret2015illuminating}, an evolutionary algorithm that uses a multi-dimensional map instead of a population to maintain its results. This map is constructed by dividing the solution space into a group of cells based on defined behavior characterstics. Any new solution found will have to evaluated for its location in the map and placed in the correct cell. In most situations, solutions in the same cell compete within the population and only the fittest individual survives. Because of the map maintaince and the cell competition, Map-Elites can guarantee a map of diverse and high quality solutions, after a finite number of iterations. These properties of Map-Elites have been used in other research topics throughout academia~\cite{alvarez2019empowering,alvarez2020interactive,khalifa2018talakat,fontaine2019mapping,charity2020mech}.

\subsection{AI Mixed Initiative in Games}
AI mixed initiative process is human and AI collaboration together on a creative task. This task can include painting~\cite{davis2015drawing,jacob2018creative}, designing shoes~\cite{bontrager2018deep}, composing music~\cite{mann2016ai}, etc. In this paper, we are focusing only on collaboration in the context of level design. Researchers have experimented with different approaches to find the best way of collaboration between human and machines. One of the earliest approaches involved using search based evolutionary algorithms to suggest content edits for the user based on a specific fitness function~\cite{smith2010tanagra,smith2010launchpad,shaker2013ropossum}. A drawback of using evolutionary algorithms is that it doesn't provide much diversity between the suggestions; which can become very limiting during collaboration if the suggestion is not ideal. Similarly, machine learning methods suffer from a similar problem, as the trained model usually suggests one solution. The advantage of using machine learning techniques is the ability to modify models using active learning to cater more to the user's personality and style~\cite{guzdial2018co,guzdial2019friend}.

\citeauthor{liapis2013sentient} has tried to solve this problem by introducing several different suggestions that improve different fitness functions during the design of 2D strategy maps~\cite{liapis2013sentient}. Unfortunately, the system was evaluating multiple different functions making it slow and hard to scale. \citeauthor{alvarez2019empowering} started using constrained Map-Elites~\cite{khalifa2018talakat} towards designing a collaborative tool to designing top down dungeons~\cite{alvarez2019empowering,alvarez2020interactive}.

\subsection{Open source projects}
Open source collaborative projects are projects that allow users to contribute specific content to create the improvement of a system. The loose structure of open source projects can discourage users from engaging with it. As such, they only join when necessary and have time. \citeauthor{brito2015towards} suggests gamifing the process in order to have more engagement with the sytem and provided a conceptual framework on how to do so in crowd sourcing projects~\cite{brito2015towards}. Several research projects gamified the process in order to guarantee user participation and collect more data~\cite{good2014cure,eiben2012increased,goncalves2014game}.

In games, this is not the case; most games are closed source. Developers can invest thousands of collective hours developing a game over the course of multiple years without sharing the data. Very few commercial developers allow users to build their own content and share ideas through their system such as Super Mario Maker (Nintendo, 2015), Skyrim (Bethesda, 2011), Time Fcuk (Edmund McMillen, 2009), etc. From a research perspective, very few projects use open sourcing to generate more game content. \citeauthor{barros2018killed}'s work~\cite{barros2018killed} is known for using crowd-sourced data such as Wikipedia to create a point-click adventure game similar to the game "Where in the World is Carmen Sandiego?" (Broderbund Software, 1985). The project was able to benefit from the large corpus of open-data to generate interesting adventure games.

\subsection{Baba is You}

Baba is You (Arvi ``Hempuli'' Teikari, 2019) is a puzzle game where players can manipulate the rules of a level and properties of the game objects through Sokoban-like movements of pushing word blocks around the map. Since the mechanics act as constraints and solutions for the player, the game provides an interesting study for procedural level generation where game levels and game rules are intertwined.

\begin{figure}
    \centering
    \begin{subfigure}[t]{0.4\linewidth}
        \centering
        \includegraphics[width=0.98\linewidth]{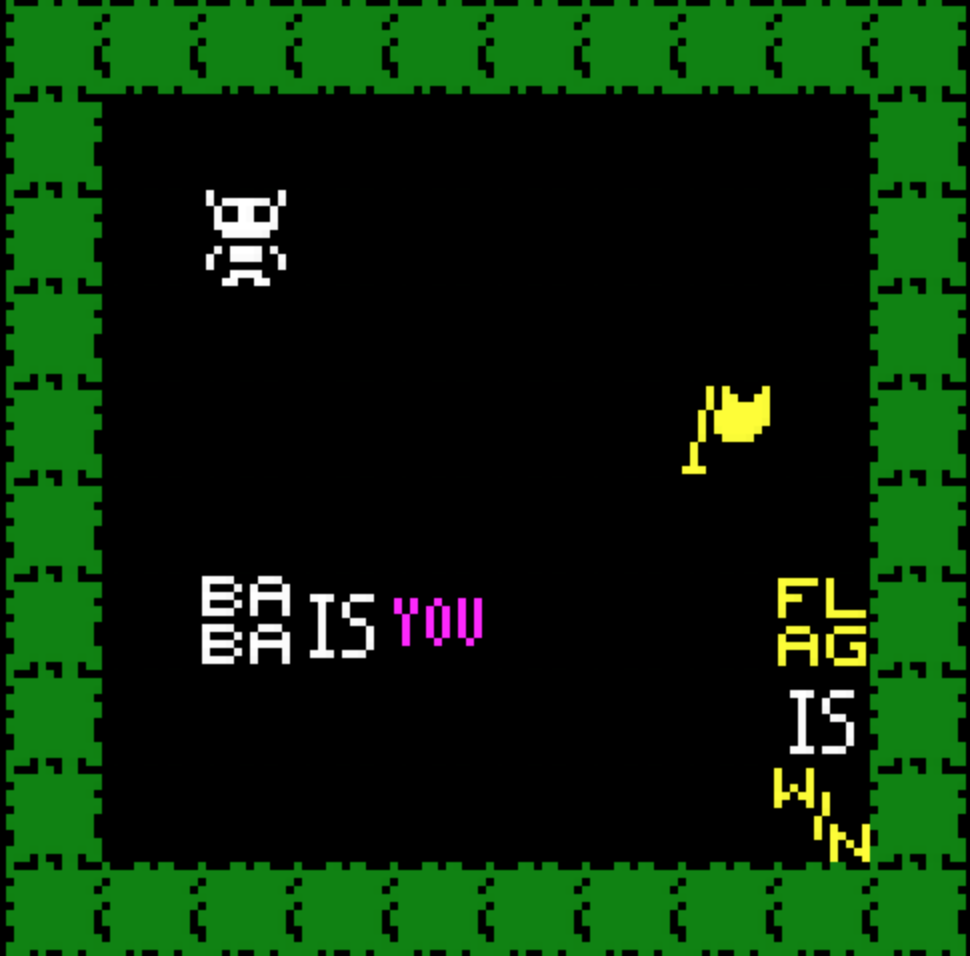}
        \caption{`Flag is Win' is satisfied at the start}
        \label{fig:baba_start}
    \end{subfigure}
    \begin{subfigure}[t]{0.4\linewidth}
        \centering
        \includegraphics[width=0.98\linewidth]{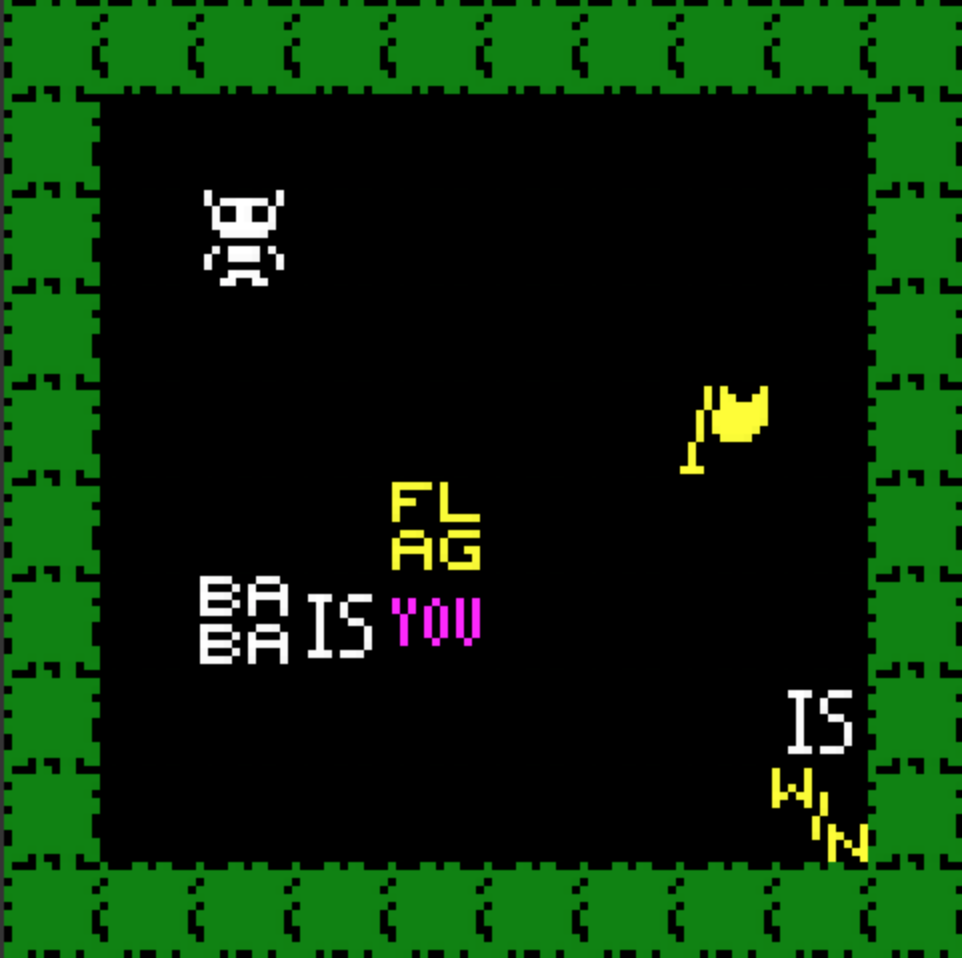}
        \caption{`Flag is Win' is not satisfied at the start}
        \label{fig:baba_notstart}
    \end{subfigure}
    \caption{Simple `Baba is You' levels with different satisfied starting rules.}
    \label{fig:babaisyou}
\end{figure}

Every `Baba is You' level must have a win condition and a controllable character that can reach the defined win condition in order to successfully solve the level. These two rules are defined by `X-IS-WIN' and `Y-IS-YOU' respectively - where X and Y are any type of game objects that can exist in the game level. Figure~\ref{fig:babaisyou} shows two simple levels from `Baba is You'. Figure~\ref{fig:baba_start} shows a simple level where the player controls the `Baba' object (with the `Baba-IS-YOU' rule is currently active) and tries to reach the `Flag' object to win the level (with the `Flag-IS-WIN' rule is active at the start of the level). The rules do not have to be initially made at the start of the game and can be manipulated at any point while the player is attempting to solve the level. Figure~\ref{fig:baba_notstart} shows the same level as figure~\ref{fig:baba_start} but the rule `Flag-IS-WIN' is not satisfied at the beginning. The player will need to push the word `Flag' to complete the sentence 'Flag-IS-WIN' - thereby activating that rule so they can win the level.

Rules for the level are defined by statements readable as English text and read from up-down and left-right. So 'X-IS-YOU' would be interpreted as a valid rule, but not 'YOU-IS-X.' As such, all rules must take one of these 3 formats:
\begin{itemize}
    \item \textbf{X-IS-(KEYWORD)} where KEYWORD belongs to a word in the keyword word class that manipulates the property of the game object class X (i.e. 'WIN', 'YOU', 'MOVE', etc.)
    \item \textbf{X-IS-X} a reflexive rule stating that the game object class 'X' cannot be changed to another game object class.
    \item \textbf{X-IS-Y} a transformative rule changing all game objects of class X into game objects of class Y.
\end{itemize}
The objects themselves ultimately do not matter in the game. For example, a player can control Baba, Keke, walls, or any other object or multiple objects at once in the game so long as it can be represented by a physical object(s) and the rule 'X-IS-YOU' is connected where X represents the physical object that the player is currently controlling. This applies to all rules created initially before the player starts to solve the level and any rules that are created or removed while the player is solving the level. Game object classes that are affected by any rule have their properties updated accordingly. For example, creating the rule 'KEKE-IS-MOVE' within the game would make all KEKE objects start autonomously moving within the game after each action is taken thereafter. If the rule is removed after a future action is taken, all KEKE objects would stop moving subsequently. 

\section{Baba is Y'all}

Baba is Y'all \footnote{http://game.engineering.nyu.edu/babaisyall/} is an Open Source Mixed initiative tool that allows users to build their own Baba is You levels and save them online. The base functionality of editing, saving and playing others' levels is similar to Super Mario Maker (Nintendo, 2015), though of course not as advanced. The system implements the jam version \footnote{https://hempuli.itch.io/baba-is-you} of Baba is You and not the more full-fledged commercial version as the jam version has enough variety in game mechanics that allows the users to build different levels. However, Baba is Y'all is not just an ordinary level editor that saves the levels online. The level archive is structured as a quality-diversity evolutionary algorithm~\cite{pugh2016quality}, specifically MAP-Elites~\cite{mouret2015illuminating}. The MAP-Elites algorithm will allow us to save, organize, retrive, suggest levels easily as they can be stored in a MAP cell corresponding to their Behavior Characteristics.

The overall process of users editing levels and building on each others' level can be seen as an instantiation of this algorithm, which seeks to improve the overall quality of the saved levels while maintaining the diversity of the levels. Every time a user submits a new level, the system saves it with mechanically similar levels (levels that use similar rules in their solutions) where they will be rated with respect to each other. Later, when a user is trying to play a new level that is mechanically different (levels that need the player to use different rules to solved them) from what they have previously play, the system can provide a different level with sufficient quality. The system also pushes the users towards designing levels that lack a specific mechanic so players can always find something new and interesting to play. Lastly, the system provides a simple evolutionary algorithm as part of the creation process to help inspire designers.

To make sure the system is simple and easy to access by everyone, we decided to design it as a web application where anyone with a broswer can access it. We modularized the system into four main modules, to make it easy to maintain and update at any time.
\begin{itemize}
    \item \textbf{Game Module:} the core game framework where the user can test game levels by solving it themselves or by using an AI solver.
    \item \textbf{Editor Module:} the level editor where the user can design Baba is You levels and submit it to our map module.
    \item \textbf{Generator Module:} provides the user with access to an evolutionary algorithm that can help them generate or modify levels. 
    \item \textbf{Map Module:} responsible for maintaining the different game levels and suggests new levels to play or make.
\end{itemize}
These modules communicate with each other based on the sequence of actions the user makes while navigating the site. 

\subsection{Game Module}
The game module is responsible for simulating a Baba is You level. It also allows users to test the playability of levels either by directly playing through the level themselves or by allowing a solver agent to attempt to solve it. This component is used in the Editor module, the Generator module, and the rating page. After a level has been saved to the database, users can attempt to solve the level in fewer moves than what was originally solved by the creator.

\begin{figure}
    \centering
    \includegraphics[width=0.6\linewidth]{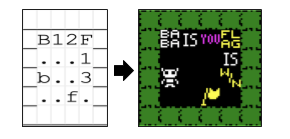}
    \caption{Example of an ascii representation of a level to a render of the level.} 
    \label{fig:ascii_map}
\end{figure}

The module receives an input level in the form of a 2D array of characters where each character represents a different game sprite. Figure~\ref{fig:ascii_map} shows an example input ascii level and its corresponding game level with different game sprites. Game sprites are divided into two main different classes: object class and keyword class. Object class are the actual game object that game rules manupilate and they are the sprites that can modify the game state, while keyword class defines the actual rules of the level. For example, figure~\ref{fig:ascii_map} shows five different keyword class sprites (BABA, IS, YOU, FLAG, and WIN). These sprites are arranged in two rules: `BABA IS YOU' allowing the player to control all the Baba objects and `FLAG IS WIN' indicating that reaching any flag object will make the player win the level. The system has a total of 32 different sprites: 11 object classes and 21 keyword classes. 

Because the game allows rule manipulation, object classes mean little in the game except for providing variety of objects for rules to affect and/or aesthetic pleasure. Because the game rules are always changing, the system keeps track at every state of all the active rules. Once the win condition has been met, the game module records the current solution, the active rules at the start of the level, and the active rules when the solution has been reached. These properties are saved to be used and interpreted by the Map module. The rules activated are used as the level's characteristic feature representation and saved as a chromosome to the MAP-Elites matrix. 

The game module provides an AI solver called 'KEKE' (based on one of the characters traditionally used as an autonomous 'NPC' in the game). KEKE uses a best-first tree search algorithm that tries to solve the input level. The branching space is based on the five possible inputs a player can do within the game: move left, move right, move up, move down, and do nothing. The algorithm uses a heuristic function based on a weighted average of the Manhattan distance to the centroid distance for 3 different groups: keyword objects, objects associated with the 'WIN' rule, and objects associated with the 'PUSH' rule. These were chosen based on their critical importance for the user solving the level - as winning objects are required to complete the level, keyword objects allow for manipulation of active rules, and pushable objects can directly and indirectly affect the layout of a level map and therefore the accessability of player objects to reach winning objects. The heuristic function is represented by the following equation:
\begin{equation}
    h = (n + w + p) / 3
\end{equation}
where $h$ is the final heuristic value for placement in the priority queue, $n$ is the minimum Manhatttan distance from any player object to the nearest winnable object, $w$ is the minimum Manhatttan distance from any player object to the nearest word sprite, and $p$ is the minimum Manhatttan distance from any player object to the nearest pushable object.

Since the system runs the solver on the user-end of the site, the solver agent has a limit of 10000 search steps to avoid computational strain to the user's machine. As such, the agent is unable to solve levels with long or complex solutions due to its limited search space. The average limit for solution steps for KEKE has not been tested yet since the focus of this project was to allow mixed-initiative collaboration with level design.

\subsection{Editor Module}

\begin{figure}
    \centering
    \includegraphics[width=0.9\linewidth]{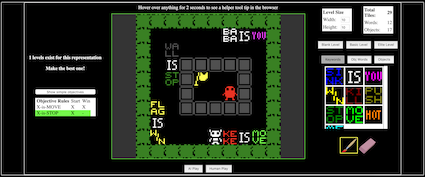}
    \caption{A screenshot of the level editor page} 
    \label{fig:editor_screen}
\end{figure}

The editor module of the system allows human users to create their own Baba is You levels in the same vain of Super Mario Maker (Nintendo, 2015). Figure~\ref{fig:editor_screen} shows the editor window that is available for the user. The user can place and erase any game sprite or keyword at any location on the map using the provided tools. As a basis, the user can start modifying either a blank map, a basic map (a map with X-IS-YOU and Y-IS-WIN rules already placed with X and Y objects), or an elite level provided by the Map Module. Similar to Mario Maker, the created levels can only be submitted after they are tested by the human player or the AI agent to check for solvability. For testing the level, the editor module sends the level information to the game module to allow the user to test it. 

Our editor also nudges the player towards creating levels with a certain mechanical goals (as seen in figure~\ref{fig:editor_screen}). The mechanical goals are defined based on the presence of certain active rules in the game either at the start state or at the winning state of the current level. The mechanical goals are written in form X-IS-KEYWORD such as X-IS-MOVE or X-IS-STOP. The X is never specified since any object in Baba is You can be used for anything while the KEYWORD is pre-defined. A list of all the different goals will be discussed later in section~\ref{map_sec}. Any combination of goals can be represented in the level - and unmade combinations will be presented to the user in order of simplicity. Through this process, the user and the system help each other to develop a large and diverse level pool. The user provides the system with novel levels, while the system guides the user by telling them what levels are not existing in the current pool to encourage originality and diversity. At any point, if the designer struggles with a concept while designing a level, they can transfer their current level to the Generator module to automatically modify the level for them. After mutating and evolving their level through the Generator, they can transfer their level back to the Editor module to do the final edits manually before submitting it to the Map module.

\subsection{Generator Module}

\begin{figure}
    \centering
    \includegraphics[width=0.4\linewidth]{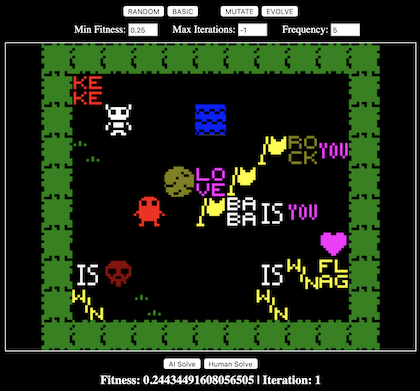}
    \caption{A screenshot of the level evolver page}
    \label{fig:evolver_screen}
\end{figure}

The Generator module is a procedural content level generator. In particularly, it is an evolutionary level generator using a fitness function based on a version of tile-pattern Kullback-Liebler Divergence (TPKLDiv\footnote{https://github.com/amidos2006/ETPKLDiv}) algorithm~\cite{lucas2019tile}. Figure~\ref{fig:evolver_screen} shows the current interface used by the evolver. As mentioned before, the Evolver module can be accessed from the Editor module at any time and can export its output back to the Editor module. With this transfer process, the generation process loses its pure procedurally generated quality. However, this process encourages mixed-initiative interaction between the algorithm and the user. This also creates an "algorithm-first" approach to creating levels, where the levels are generated towards a fitness value and then edited by the user later. The evolver interface provides the user with multiple customizations such as the initialization method, stopping criteria, evolution pausing, and an application of a mutation function allowing manual user control. With these features, the user is not directly changing the evolution process itself, but instead guiding and limiting the algorithm towards generating the level they want.

The original ETPKLDiv algorithm uses a 1+1 evolution strategy, a.k.a. a hillclimber, to improve the similarity between the current evolved chromosomes and a reference level. The algorithm uses a sliding window of a fixed size to calculate the probability of each tile configuration (called tile patterns) in both the reference level and the evolved level and tries to minimize the Kullback-Liebler Divergence between them. For this project, we used a window size of 3x3. This was to maximize the probability of generating initial rules for a level, since rules in Baba is You are made up of 3 tiles.

In our project, we used 2+2 evolution strategy instead of 1+1 used by \citeauthor{lucas2019tile} to allow slightly more diversity in the population. We also modified the fitness function to allow it to compare with more than one level. The fitness value also includes the playability of the level ($p$), the ratio of empty tiles to total level tiles ($s$), and the ratio of unnecessary objects ($u$). The final fitness equation for a level is as follows:
\begin{equation}
    fitness_{new} = min(fitness_{old}) + u + p + 0.1 \cdot s
\end{equation}
where $fitness_{old}$ is the Kullback-Lievler Divergence fitness function from \citeauthor{lucas2019tile}'s work~\cite{lucas2019tile} compared to the most similar reference levels.

The value $u$ is the percentage of unnecessary objects in the map; objects that are not required or predicted to act as a constraint or solution for the level. The value can be calculated as follow:
\begin{equation}
    u = \frac{o}{a}
\end{equation}
where $o$ is the number of objects sprites initialized in the level without a related object-word sprite and $a$ is the total number of object sprites initialized in the level. This value is implemented in order to prevent noise within the level due to having object tiles that cannot be manipulated in any way or have relevancy to the level. A human-made level may include these "useless" tiles for aesthetic purposes or to give the level a theme - similar to the original 'Baba is You' levels. However, the PCG algorithm optimizes towards efficiency and minimalist levels, therefore ignoring the subjective aspect of a level's quality (which can be added later by the user).

The value $p$ is a binary constraint value that determines whether a level is potentially winnable or not. The value can be calculated as follows:
\begin{equation}
    p = 
    \begin{cases}
        1, & \text{has ['X-IS-YOU' rule, 'WIN' keyword]} \\
        0, & otherwise
    \end{cases}
\end{equation}
This is to ensure any levels that are absolutely impossible to play or win are penalized in the population and less likely to be mutated and evolved from in future generations. We used a simple playability constraint check instead of checking for playability using the solver because the solver take time to check for playability. Also all playable levels by the solver usually end up being easy levels due to the limited search space we are given for the best first algorithm.

The value $s$ is the ratio of empty space tiles to all of the tiles in the level. The equation can be calculated as follows:
\begin{equation}
    s = \frac{e}{t}
\end{equation}
where $e$ is the number of empty spaces in the level and $t$ is the total number of tiles found in the level. The value $s$ is multiplied with a value of 0.1 to avoid heavy penalization for having any empty spaces in a level and to prevent encouragement for levels to mutate towards populating the level with an overabundance of similar tiles in order to eliminate any empty space. This instead causes the algorithm to favor smaller and more compact levels over levels with a large area.

The Generator module is not run as a back-end process to find more levels, instead it has to be done manually by the user. This is done due to the fact that some generated levels cannot be solved without human input. One might wonder why not generate a huge corpus of levels and ask the users later to test them for the system. This could result in the system generating a multitude of levels that are either impossible to solve or are solvable but not subjectively "good" levels - levels the user would not find pleasing or enjoyable. This overabundance of "garbage" levels could lead to a waste of memory and a waste human resources. By allowing the user direct control over which levels are submitted from the generation algorithm, it still guarantees that the levels are solable and with sufficient quality and promote using the tool in a mixed-initiative approach.

\subsection{Map module}
\label{map_sec}

The Map module is the core module of the system. It provides three main functionalities: level archiving, suggesting levels to create, and suggesting levels to be played and rated. In order to achieve all of these functionalities, we need an optimal way of representing the relationship between the created levels. A way to measure similarity between levels so we can find the best levels between them and easily discover new levels every time. We ended up using the same structure as the Map-Elites algorithm~\cite{mouret2015illuminating} in order to guarantee quality and diversity between the levels. 

\begin{table}[t]
    \caption{Chromosome Rule Representation}
    \centering
    \begin{tabular}{|p{0.2\linewidth}|p{0.7\linewidth}|}
    \hline
         Rule Type & Definition \\
    \hline
    \hline
        X-IS-X & objects of class X cannot be changed to another class \\
        X-IS-Y & objects of class X will transform to class Y \\
        X-IS-PUSH & X can be pushed \\
        X-IS-MOVE & X will autonomously move \\
        X-IS-STOP & X will prevent the player from passing through it\\
        X-IS-KILL & X will kill the player on contact\\
        X-IS-SINK & X will destroy any object on contact\\
        X-IS-[PAIR] & both rules 'X-IS-HOT' and 'X-IS-MELT' are present \\
        X,Y-IS-YOU & two distinct objects classes are controlled by the player \\
    \hline
    \end{tabular}
    \label{rrp}
\end{table}

When a level is submitted to be archived, the system uses the the list of active rules at the start and the end of the level as behavior characteristic for the input level to determine its location in the map. There are 9 different rules checked for in each level - based on the possible rule mechanics that can be made in the Game module system. Table \ref{rrp} shows the full list of possible rules. Since these rules can be active at the beginning or at the end, it makes the number of behavior characteristics equal to 18 instead of 9 which proivde us with a map of $2^{18}$ cells.

In this project we are using a multi population per each cell of the Map-Elites similar to the constrained Map-Elites~\cite{khalifa2018talakat}. In our project, we are using two tables/populations per cell named: rated table and elite table. The 'Rated' table contains all the levels that are created either by humans, the system, or both. On the other hand, the 'Elite' table only contains one level. This level must have an average rating score of at least 3.5 out of 5 based on 5 different ratings. The elite can be replaced if a new level appears with higher average rating than the previous one.

A rating for a single level is determined by comparison to another level within the entire 'Rated' table population. The user must determine the better level based on two qualities: hardness and design. A level that is considered 'harder' represents that the solution search space for the level takes longer to arrive at - and therefore is more challenging to the user. A level that is considered to have 'better design' represents that the level is more visually pleasing and elegant with its map representation - a quality that is hard to generate automatically with AI. The minimum score rating for a level is 1 and adds 2 for each comparison category it wins - resulting in a maximum score of 5. 

\begin{figure}
    \centering
    \includegraphics[width=0.8\linewidth]{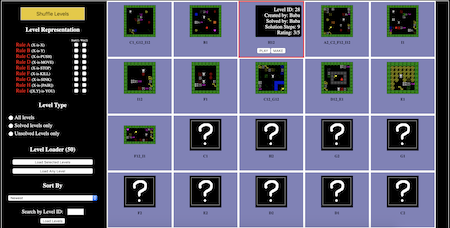}
    \caption{A screenshot of the level matrix page}
    \label{fig:levelMat}
\end{figure}

The user at anytime can check the state of the MAP-Elites population by visiting the level matrix page. The level matrix shows 50 randomly chosen cells that can be filtered and sorted according to a certain behavior characteristic. Figure~\ref{fig:levelMat} shows the level map interface provided to the user. The retrived levels are sorted from simplest to hardest and the set can be reshuffled to retrieve new levels at anytime. From here, the user can choose between playing a level that has already been made to find a faster solution, editing a submitted level to either create a better level for a MAP-Elites cell population or contribute to a different cell's population by altering the rules, or selecting an unpopulated cell to make a new novel level representation. Our main intention is to encourage users to create new levels and contribute to filling the cells of the MAP-Elites matrix.

\section{Study}

The study results were collected 32 hours after the announcement release of the system. Out of the $2^{18}$ cells in the MAP-Elites matrix, 38 had at least one level. 58 levels were made in total - 12 were user-only created levels, 11 were evolver-only created levels, and 35 were evolver-user collaboration levels. 18 users had accounts registered on the site. The average number of solution steps for the levels was 27.93 steps. 

\begin{figure}
    \centering
    \includegraphics[width=0.9\linewidth]{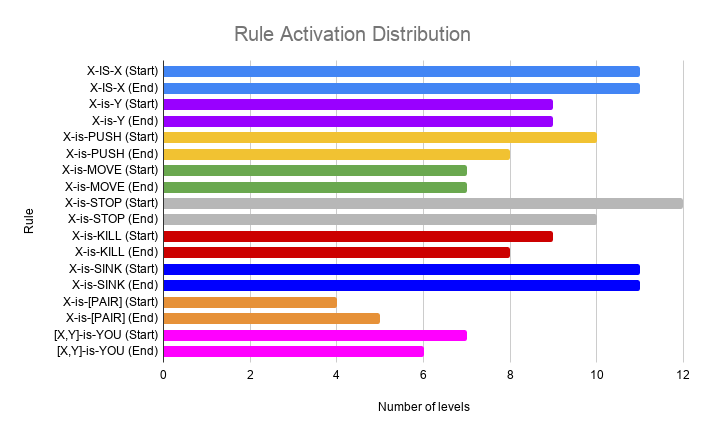}
    \caption{Distribution histogram of the rules activated in the levels created}
    \label{fig:ruledist}
\end{figure}

\begin{figure}
    \centering
    \begin{subfigure}[t]{.3\linewidth}
        \includegraphics[width=0.95\linewidth]{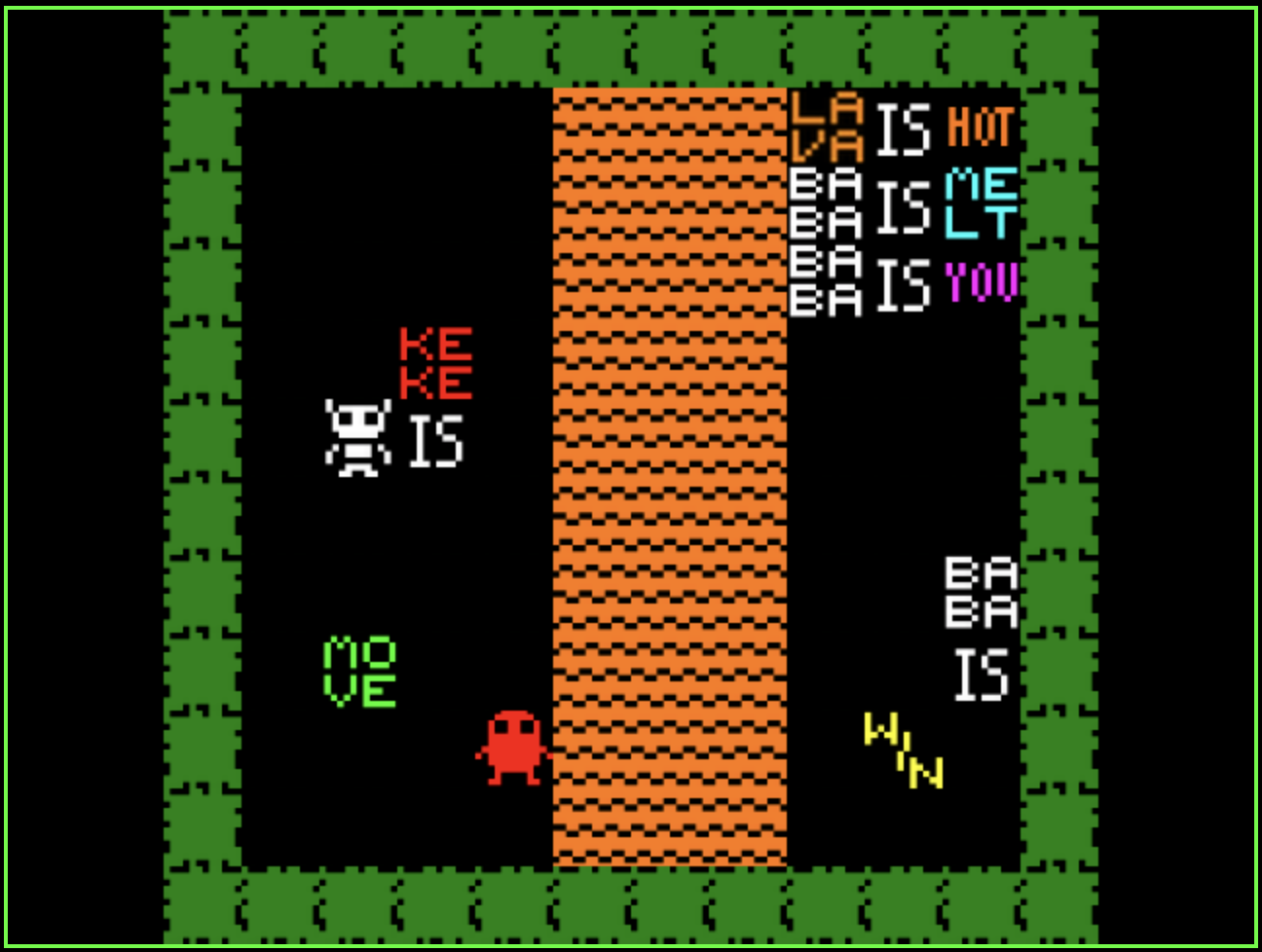}
        \label{fig:user1}
    \end{subfigure}
    \begin{subfigure}[t]{.3\linewidth}
        \includegraphics[width=0.95\linewidth]{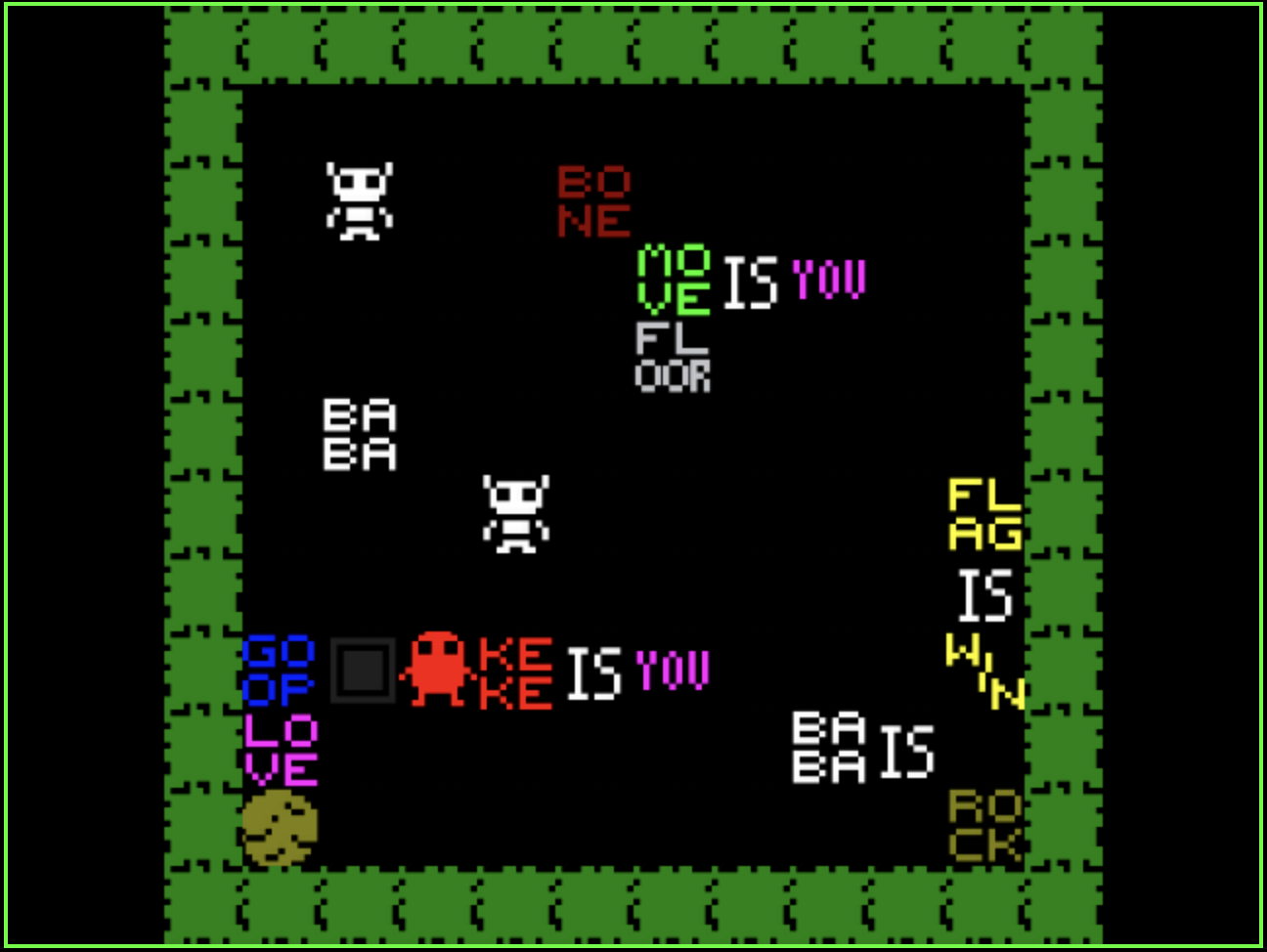}
        \label{fig:pcg1}
    \end{subfigure}
    \begin{subfigure}[t]{.3\linewidth}
        \includegraphics[width=0.95\linewidth]{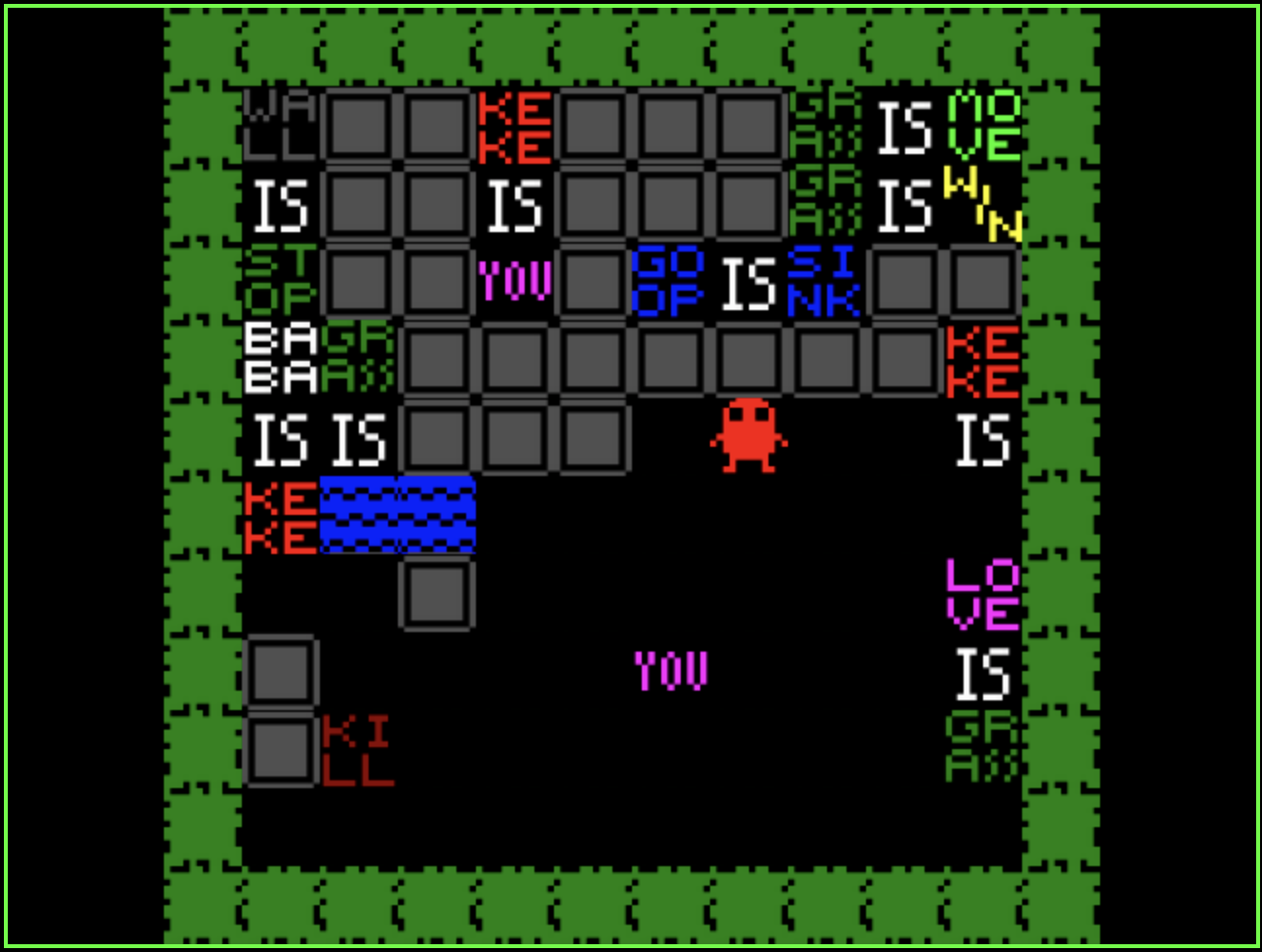}
        \label{fig:pcguser1}
    \end{subfigure}
    \begin{subfigure}[t]{.3\linewidth}
        \includegraphics[width=0.95\linewidth]{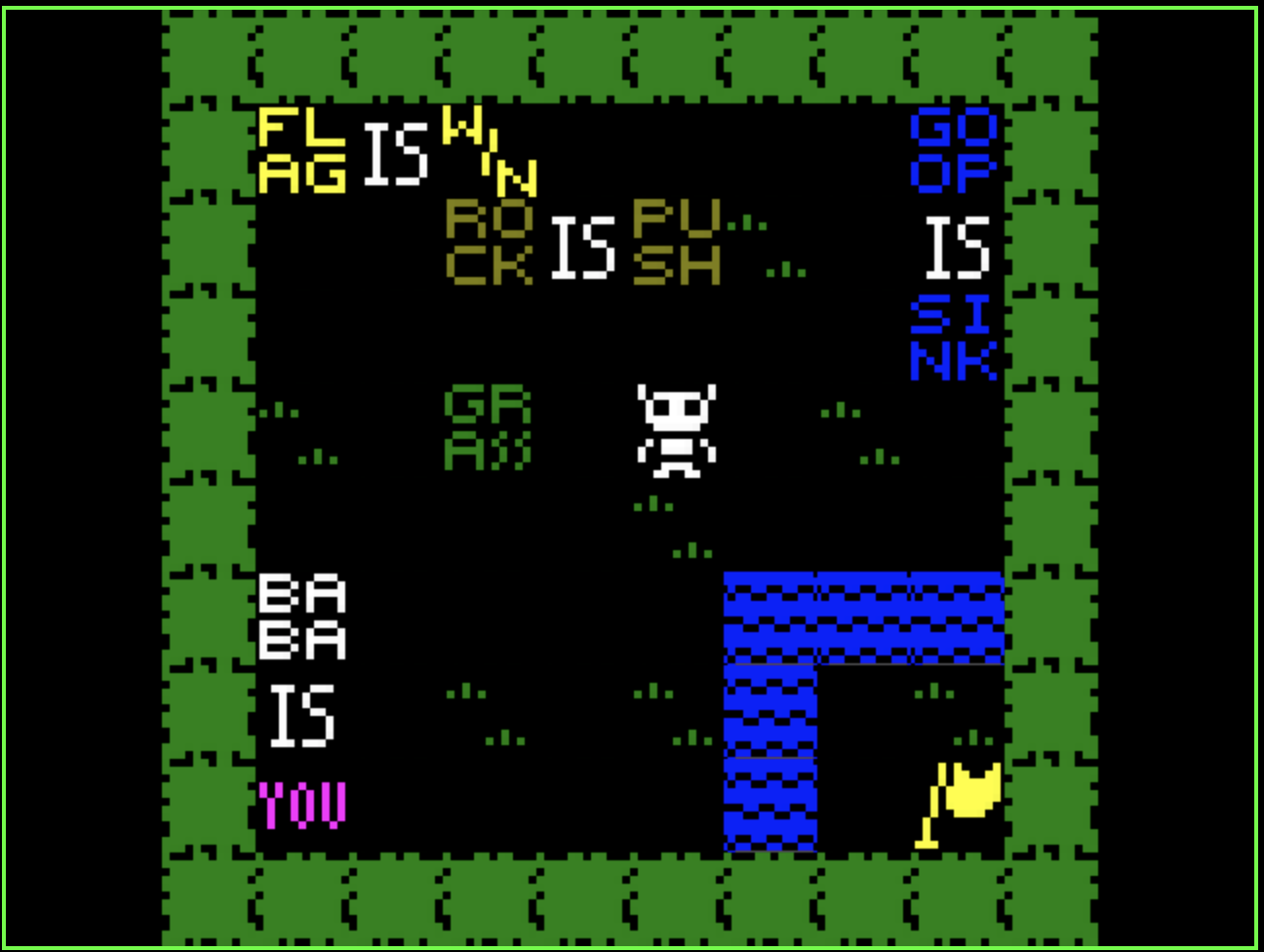}
        \label{fig:user2}
    \end{subfigure}
    \begin{subfigure}[t]{.3\linewidth}
        \includegraphics[width=0.95\linewidth]{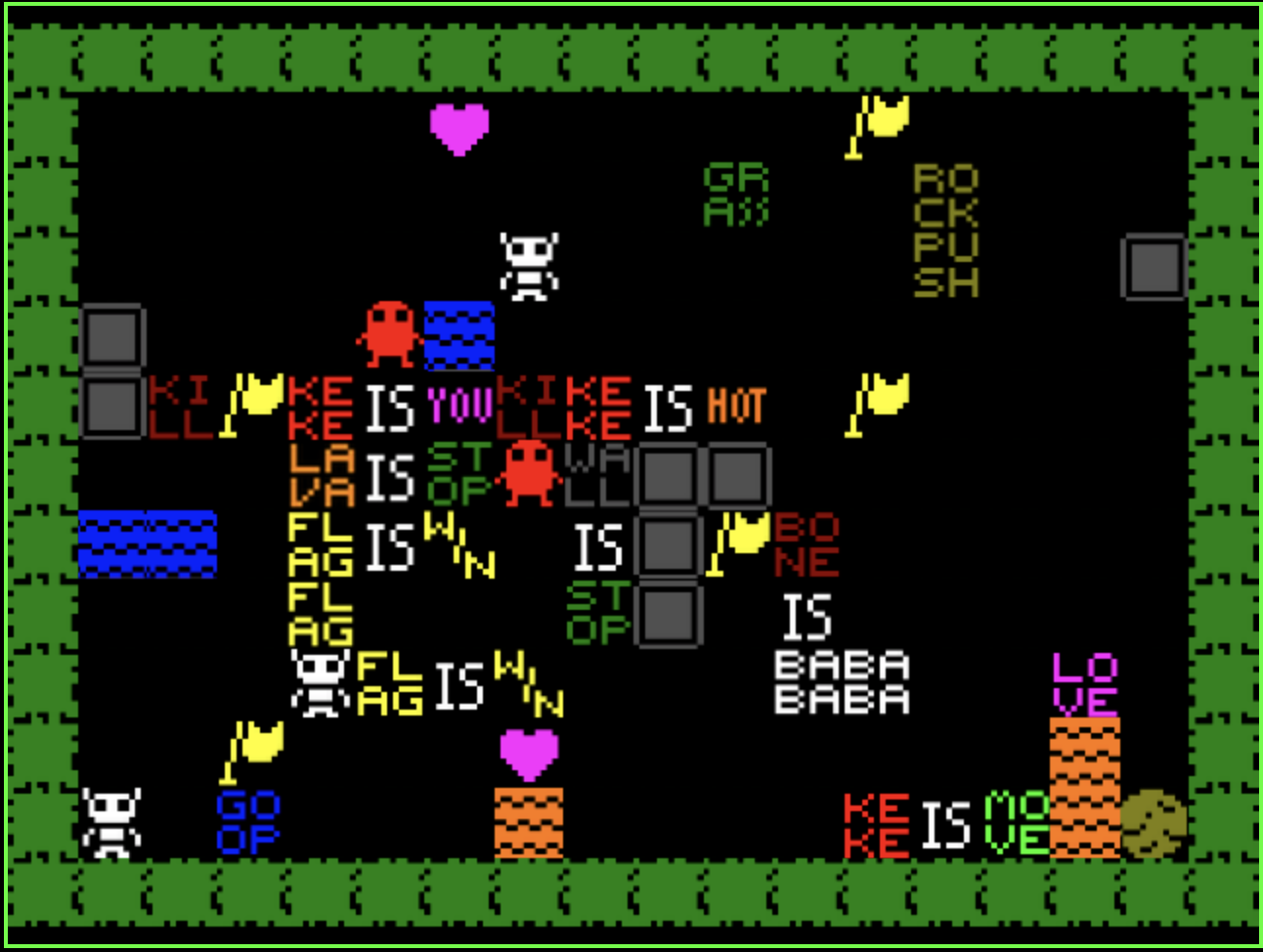}
        \label{fig:pcg2}
    \end{subfigure}
    \begin{subfigure}[t]{.3\linewidth}
        \includegraphics[width=0.95\linewidth]{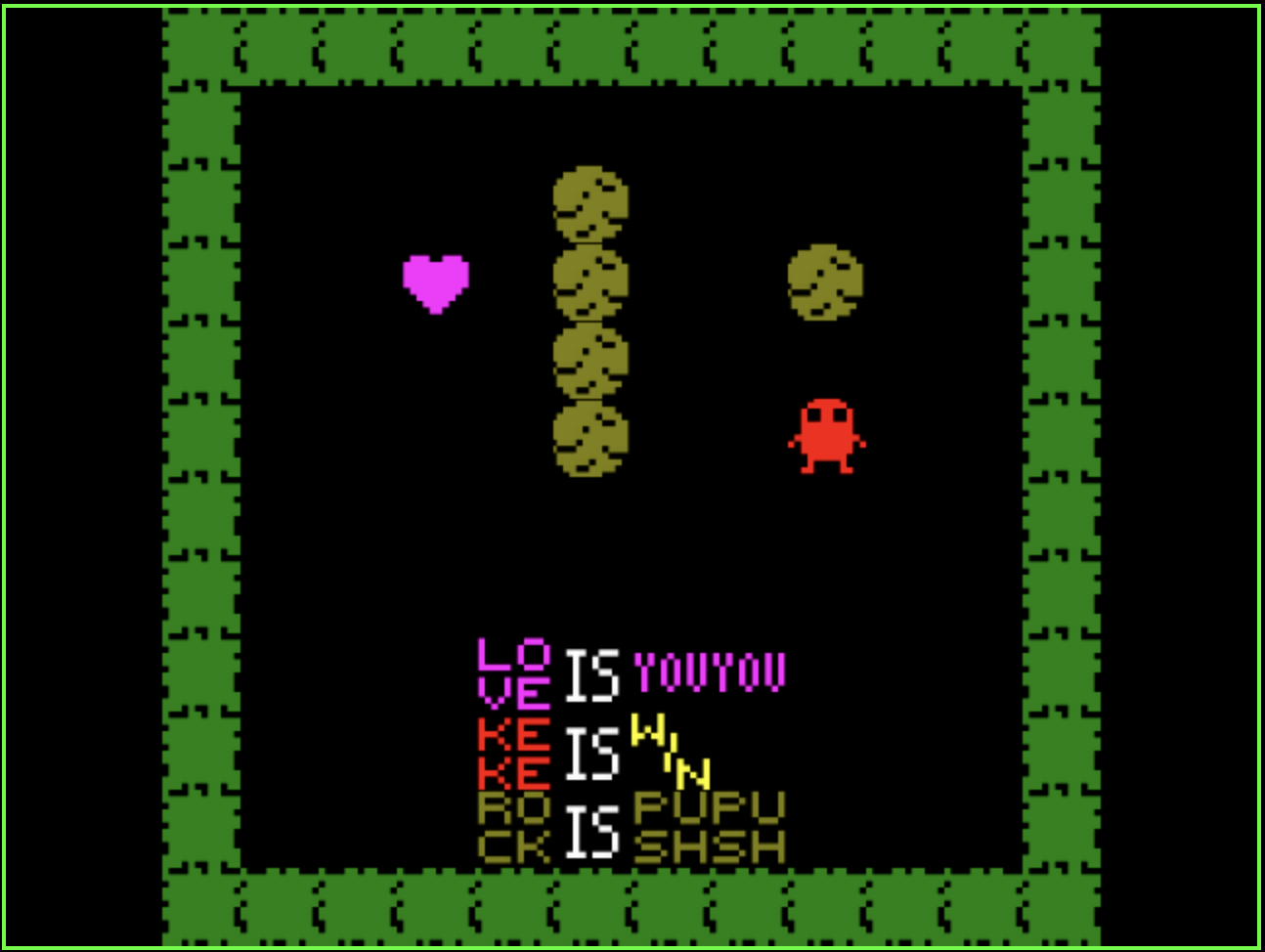}
        \label{fig:pcg3}
    \end{subfigure}
    \vspace{0.2cm}
    \begin{subfigure}[t]{.3\linewidth}
        \includegraphics[width=0.95\linewidth]{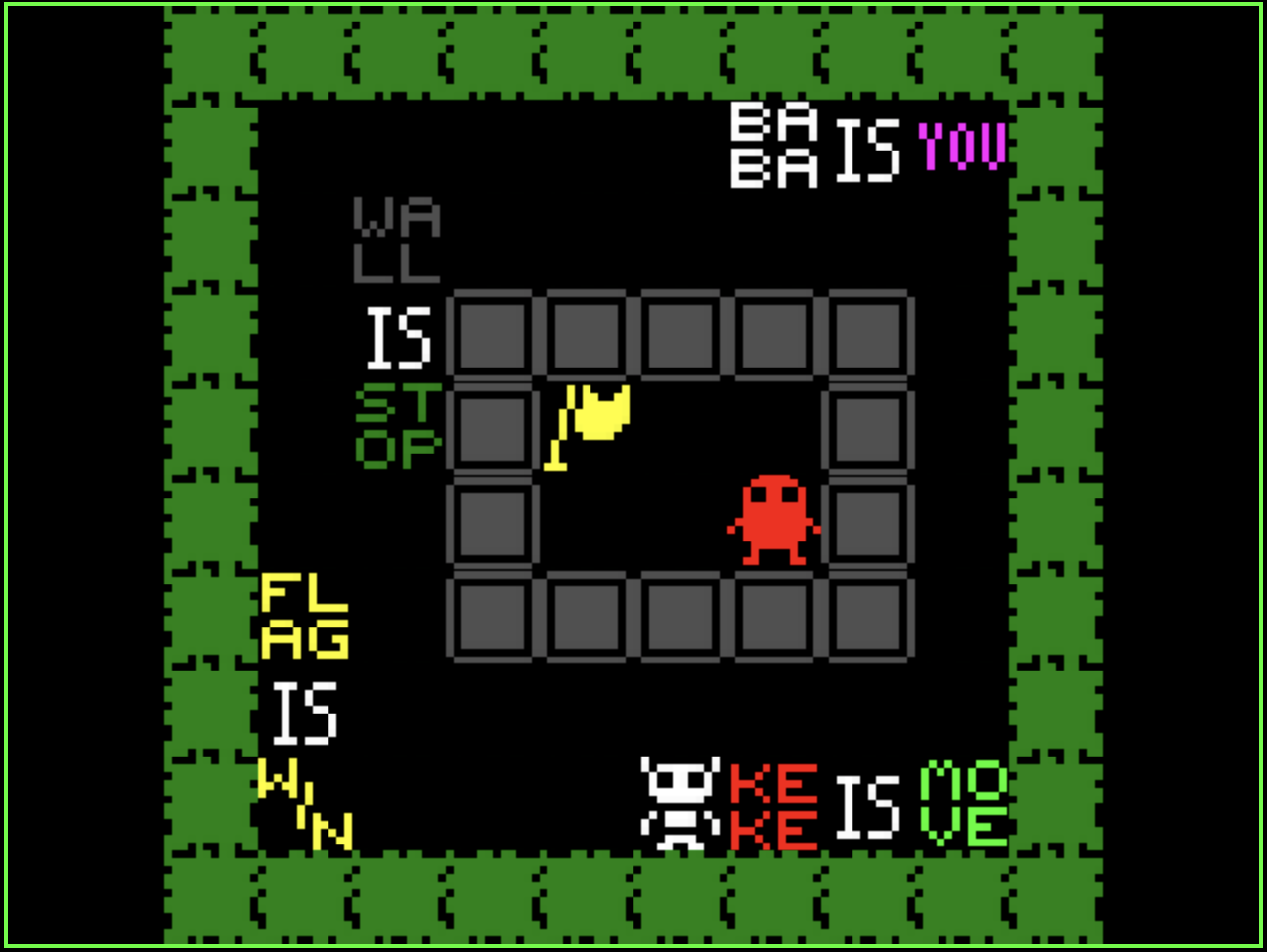}
        \label{fig:user3}
    \end{subfigure}
    \begin{subfigure}[t]{.3\linewidth}
        \includegraphics[width=0.95\linewidth]{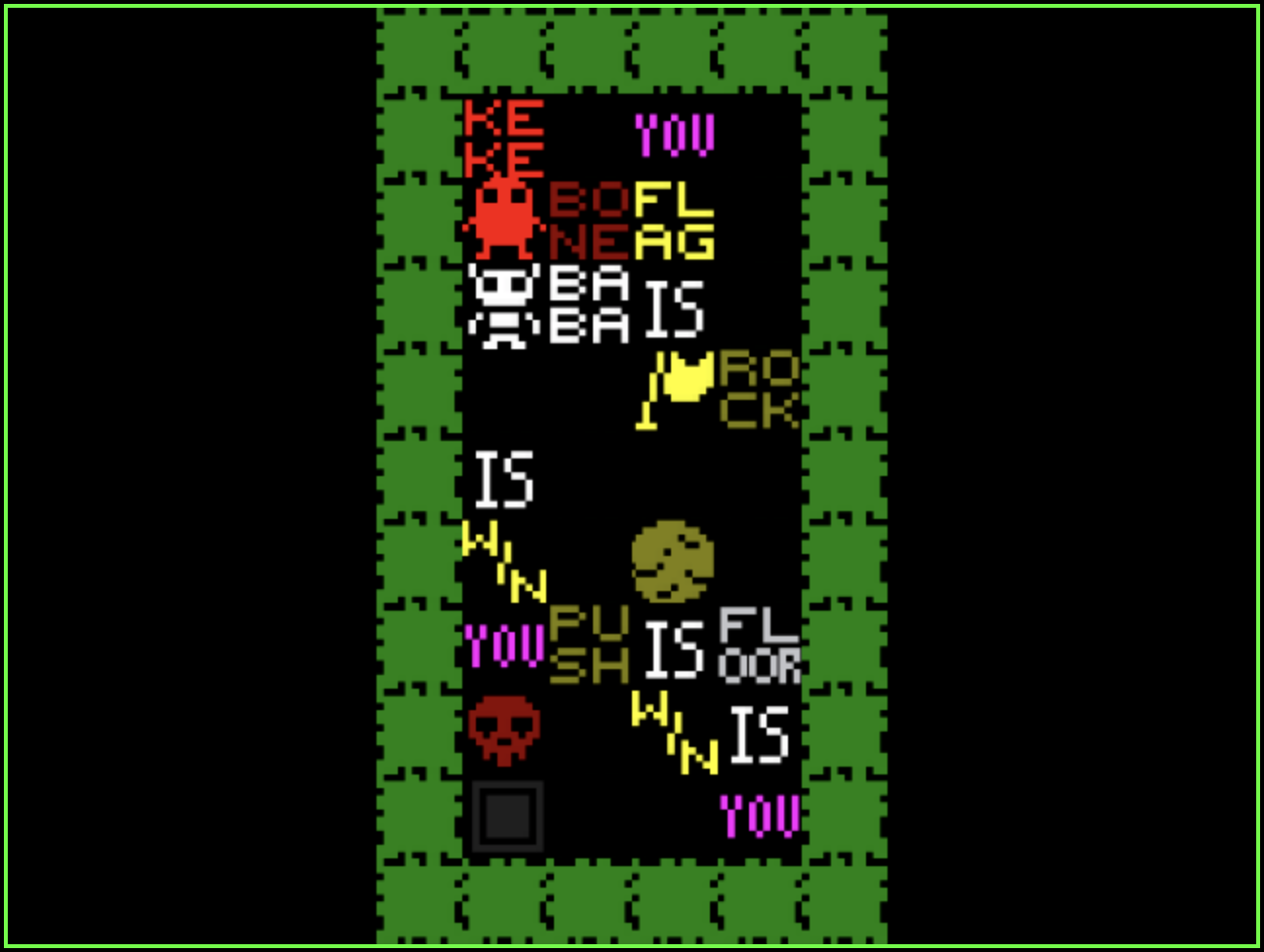}
        \label{fig:pcg3}
    \end{subfigure}
    \begin{subfigure}[t]{.3\linewidth}
        \includegraphics[width=0.95\linewidth]{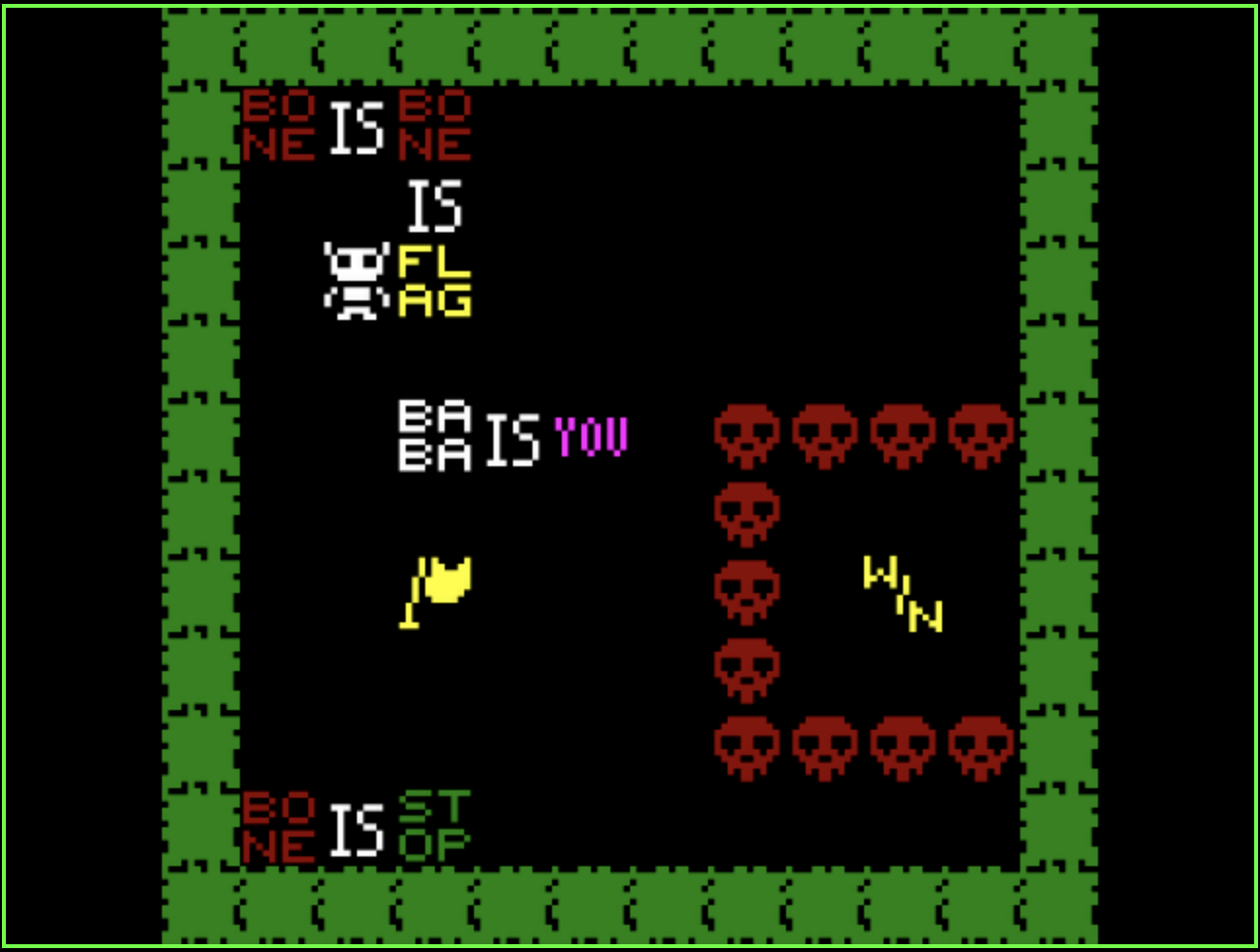}
        \label{fig:pcguser3}
    \end{subfigure}
    \caption{Sample levels generated for the system. The left column is user generated levels, the middle column is evolver module levels, and the right column is mixed-initiative user and evolver levels}
    \label{fig:sample_levels}
\end{figure}

An average of 2.924 rules were used in the level creation. A full distribution of the rules activated in each level can be shown in Figure \ref{fig:ruledist}. By looking at the figure, we can see even with the small amount of participants we had, the system was able to cover all the different mechanics that the game can offer. We can't guarantee that this high coverage was due to the system provided goals during the level creation, but it is a good indication that the system is working and helping the users to explore new level ideas. Figure~\ref{fig:sample_levels} shows some examples of the different levels submitted to the sytem.

A survey was also made available to participants through the site. We recieved two responses. One participant used the evolve tool, one participant made an account on the site, and no participants rated levels. One participant mentioned as an additional comment that they were unsure how to save a level they made.

\section{Discussion}




In the time span of data collection for this paper, the MAP-Elites matrix was 0.01\% filled. However, there is also a lack of a strong incentive for creating the levels, and the the learning curve for interacting with the system is steeper than it should be. The amount of unmade levels and prompts for the user to participate in could also be overwhelming. Regardless, the levels generated were fairly distributed with the rules - giving cause to believe that the level prompts given by the system were at least attempted by the userbase. 

Typically with crowd-sourced tasks, users participating are given an incentive to complete the task. Whether it be small monetary incentives, social recognition within the community, or personal benefit from contributing to a bigger cause, users need some reason to participate. We hypothesized that users would participate in this study based on the reason of contribution to a bigger cause and contributing to a large pool of levels along with other users and evolutionary AI. However, the user may have been overwhelmed with how many tasks were needed to be completed. It might therefore make sense to add elements of gamification to the system to motivate users.

Baba is Y'all's target audience was aimed towards users who are natural problem solvers, creative, and interested in puzzle games and content generation. This group of people would be the most motivated to contribute to task-oriented generation of levels - and thus be able to populate the cells of the MAP-Elites matrix. Other users who find themselves outside of this niche can still contribute to this generating process by evaluating levels and providing feedback to the creators and the system itself - improving the quality of the level creation overall. The end result is a full map of levels of "good" quality - made by both the creators and the evaluators.

\section{Conclusion and Future Work}

Baba is Y'all is an ongoing level creation and generation prototype system for creating Baba is You levels based on rule representations using collaborative mixed-initiative techniques between human users and evolutionary algorithms. The system prompts the user to create levels based on what it is missing in its current level set so as to create a exhaustive library of levels with specific mechanic behaviors. While there are still many more levels to be made in the creation space, the application serves as a proof of concept for task-oriented level design aided, guided, and organized by an AI system.

The project is the beginning of many more explorations with this system and collaborative mixed-initiative level design as a concept. Ideally, the individual modules of the Baba is Y'all system can be swapped out to incorporate other variations of generation algorithms, level editors, and/or mapping organization algorithms. We would like to expand this system to use other popular games for collaborative level design. Some examples include designing Zelda-like dungeons needing specific mechanics or tool combinations in order to be completed or designing platforming levels that require specific combinations of actions to reach the goal. We would also like to experiment with altering level space organization, level editing user interface, and different procedural content generation and evolutionary algorithms. For example, it would be useful if the designer could change the fitness function, with a suggested fitness function being to satisfy the mechanics suggested by the system.

To speed up the level generation process, we would like to run a back-end system parallel to the web-based system that can quickly generate, solve, and upload new novel levels to fill in the MAP-Elites cells. In this project, a level was only submitted if and only if the level was provided with a solution. For a user-excluded generator, only a solver agent could be used. The levels would also need to include a diversity metric and a dynamically updating input source from the user created levels in order to maintain visually appealing but solvable levels that satisfy filling the population of cells in the MAP-Elites matrix. 

For user-end features we would like to examine collaborations between multiple users and multiple types of evolutionary algorithms all at once to create levels. This would broaden the scope and possibilities of level design and development even further to allow more creativity and evolutionary progress within the system. We would also allow the option for users to personally save their levels to their account and search for other user-specific levels and share their levels with other users. This will hopefully make the site more engaging for the user - with less prioritization on population of cells as was originally designed. In addition to extending user features and interactions, we plan to focus on looking into more human-computer interaction-based resources to make the site and system more user-friendly. The new site design will be evaluated by a user study to gain feedback and make continuous improvements to layout and formatting - like any website with a large userbase. This will hopefully encourage more mixed initiative interaction between the human and the system and thus create more creative and innovative levels.

\section*{Acknowledgments}
Ahmed Khalifa acknowledges the financial support from the NSF Award number 1717324 - ``RI: Small: General Intelligence through Algorithm Invention and Selection.''.

\bibliographystyle{IEEEtranN}
\bibliography{IEEEabrv,IEEEbiblo}

\end{document}